%% file: ICC19.tex
\documentclass[conference]{IEEEtran}
\IEEEoverridecommandlockouts
\usepackage{amsmath,amssymb,amsfonts}
\usepackage{algorithmic}
\usepackage{graphicx}
\usepackage{textcomp}
\usepackage{xcolor}
\usepackage{makecell}
\usepackage{epstopdf}
\usepackage{subfigure}
\usepackage{makecell}

\begin{document}

\title{Latency Bounds of Packet-Based Fronthaul for Cloud-RAN with Functionality Split}

\author{\IEEEauthorblockN{Ghizlane Mountaser, Maliheh Mahlouji, Toktam Mahmoodi}
	\IEEEauthorblockA{Centre for Telecommunications Research \\ Department of Informatics, King's College London \\ London WC2B 4BG, UK} \\
	\{ghizlane.mountaser, maliheh.mahlouji, toktam.mahmoodi\}@kcl.ac.uk}

\maketitle

\begin{abstract}
The emerging Cloud-RAN architecture within the fifth generation (5G) of wireless networks plays a vital role in enabling higher flexibility and granularity. On the other hand, Cloud-RAN architecture introduces an additional link between the central, cloudified unit and the distributed radio unit, namely fronthaul (FH). Therefore, the foreseen reliability and latency for 5G services should also be provisioned over the FH link. In this paper, focusing on Ethernet as FH, we present a reliable packet-based FH communication and demonstrate the upper and lower bounds of latency that can be offered. These bounds yield insights into the trade-off between reliability and latency, and enable the architecture design through choice of splitting point, focusing on high layer split between PDCP and RLC and low layer split between MAC and PHY, under different FH bandwidth and traffic properties. Presented model is then analyzed both numerically and through simulation, with two classes of 5G services that are ultra reliable low latency (URLL) and enhanced mobile broadband (eMBB).


\end{abstract}

\begin{IEEEkeywords}
Cloud-RAN; Fronthaul; Ethernet; Latency; Reliability; Upper bound; Lower bound.
\end{IEEEkeywords}

\section{Introduction}

\input{Intro}

\section{Background}

\input{Background}

\section{System Model}

\input{system}

\section{Stochastic Delay Bounds for (n,k) fork-join System}

\input{bound}

\section{Simulation Results}

\input{simulation}

\section{Conclusion and Future Research}

In this paper, we presented a Cloud-RAN model based on multi-path FH with coding solution for enhancing the reliability of the FH. The paper aims at providing an upper and a lower bounds of reliability-latency function on the FH under orthogonal FH allocation.

We first derived lower and upper bounds analytically. Then we simulated the Cloud-RAN model to demonstrate the effectiveness of the analytic by showing the simulation results are bounded by lower and upper bounds. Finally, based on this result, we discussed the recommendations for split point focussing on MAC-PHY and PDCP-RLC splits for different scenarios to meet their latency and reliability requirements.

Future work will be focusing on analyzing multi-path FH with multi-hops.

\section*{Acknowledgement}
\begin{small}
This work has been supported by The Engineering and Physical Sciences Research Council (EPSRC) industrial Cooperative Awards in Science \& Technology (iCASE) award and by the British Telecom (BT). Additional support is received from EU H2020 5GCAR.
\end{small}

\bibliographystyle{ieeetr}
\bibliography{ICC19}

\end{document}

%% file: Intro.tex
\label{sec:Intro}

One of the architecture enablers of fifth generation (5G) is Cloud-RAN that is supported through multiple technological advances in the network including softwarization, virtualization and cloudification \cite{mc-nfv-sdn}. Cloud-RAN despite bringing higher flexibility and granularity to the network architecture, introduces an additional communication link, i.e. fronthaul (FH). Hence, the ability for the FH to flexibly scale up with data rate has become critical to the success of Cloud-RAN.
The need for flexibility in the FH has opened up the possibility of flexibly splitting Radio Access Network (RAN) functionalities between central unit (CU) and distributed unit (DU). The advantage to such an architectural approach is the use of different transport such as packet-based FH. Adopting packet-based FH in the Cloud-RAN architecture allows the use of widely deployed Ethernet-based network. At the same time, packet-based networks impose challenges in ensuring the high reliability and low-latency over the FH communication, which are the key performance indicators expected of 5G.

Reliability can be typically achieved by retransmission or redundancy. However, reliability is usually increased at the cost of latency which poses a major challenges when latency requirement is very stringent. For this reason, the design of Cloud-RAN solution involves one key design question, that is, which functional splits may be suitable from a reliability-latency point of view under the constraint required by 5G scenarios.


Given that each split comes with its own delay requirements, having the knowledge of latency bounds of FH will allow us to decide which split is the most appropriate in the Cloud-RAN architecture. On the other hand, 5G traffic classes eMBB (enhanced mobile broadband), URLLC (ultra reliable low latency communications) and Massive machine type communications, each come with varied requirements on reliability that should also be maintained on the FH link; improving reliability often results in increasing latency. To this end, focusing on reliable packet-based FH \cite{Ghizi}, the aim of this paper is to compute the lower bound and upper bound of latency analytically, using stochastic network calculus \cite{Fidler2015, Wang2017, Gauri}. We simulate the reliable packet-based FH, demonstrating where and how delay bounds are achieved for two classes of traffic, which are eMBB and URLLC. Having these bounds, we further analyse where each functionality split can be the best architectural choice. This paper is an expended work to our previous work \cite{Ghizi} where we investigated how to improve reliability and latency of packet based fronthauling by means of multi-path diversity and erasure coding; and \cite{7925770, 8269135} which examined lower layer and higher layer splits through an experimental testbed considering an Ethernet-based FH.

The remainder of this paper is organized as follows. Section II gives a brief overview on Cloud-RAN and its different transport technologies and explores reliability on the FH. In section III, we elaborate system model of Cloud-RAN with multi-path FH using coding to analyse reliability-latency and we shed light on functional split requirements in term of latency. In section IV, we compute stochastic delay lower and upper bounds of the system model. The analytic and simulation results are studied in section V. Finally, conclusion and future research are presented in section VI.


%% file: Background.tex
\label{sec:Background}

Cloud-RAN is considered one of the key enablers of 5G architecture, given its desired properties \cite{TheRoadTowardsGreenRan}. Despite the attractive advantages of the conventional Cloud-RAN, the architecture whereby a standard common public radio interface (CPRI) is used to transport base band radio samples between CU and DU faces several challenges. The first challenge is the user data is transmitted in the form of an IQ-data block which requires large bandwidth of $157.3$ Gbps, considering a 100 MHz transmission bandwidth and 32 antenna ports \cite{3gpp.38.801}. Thereby, the high-throughput requirement poses challenges for the FH interface. The second challenge is that CPRI requires stringent latency and jitter requirements. It requires an end-to-end latency of around $250$ $\mu$s \cite{NGMN}. Such critical requirements are making CPRI very challenging. To relax the excessive bandwidth and latency requirements, as well as to enhance the flexibility of the FH, functional split is introduced whereby a more flexible placement of baseband functionality between the DU and the CU is considered. In fact several options of functional splits have been standardized in 3GPP \cite{3gpp.38.801}. Moreover, fundamental simulations and experimentations are carried out by various academic studies \cite{8252881, 8269135, 8419201}. Nonetheless, each split point has different requirements such as latency and bandwidth. These requirements have to be considered in order to select the appropriate functional split. In general, the lower the split point, the greater the level of centralization, the higher is the required interface data rate and the more stringent is the latency requirement.

In traditional Cloud-RAN architecture, fibre has been defined as an ideal attractive solution to meet the strict requirements of high bandwidth and low latency of CPRI. However, there are situations where deployment of fibre is difficult or not a good choice due to cost. In this end, packet-based FH can be considered as a promising alternative transport. This highly cost effective solution allows sharing and convergence with Ethernet-based fixed networks and offers great flexibility. However, packet-based FH imposes many challenges such as high latency and high jitter. Nevertheless, it can be used in functional split where the latency and jitter requirements are relaxed. For example, some of the authors of this work has demonstrated the feasibility of splitting between MAC and PHY in their past work \cite{7925770}. Further studies have shown that the requirements of different 5G service classes, including the URLLC service can be accommodated using packet-based FH  \cite{8269135}. In \cite{7841885} authors analysed impact of packetization on the Cloud-RAN and they analyzed different packet scheduling to increase the multiplexing gain.

In addition to low latency and excessive data rate, reliability is also an important metric in 5G \cite{8329620,8292238}. Under the Cloud-RAN architecture, FH needs to provide comparative reliability to enable adoption of Cloud-RAN. The most well used methods to improve reliability are retransmission, multi-path with packet duplication and multi-path with coding. Retransmission is a straight forward way to achieve reliability. However, retransmission can have significant impact on increasing the latency making it non-viable solution on the FH where delays cannot be afforded. By contrast, path diversity with duplication offers better latency in the expense of significant transmission overhead by duplicating packets over multiple interfaces increasing FH network congestion. Two important considerations in such approaches are latency and FH overhead. An alternative solution that provides trade-off between latency and FH overhead is channel coding which can add controlled redundancy to achieve desired reliability and splits the total amount of information to transmit across different paths. Additional reliability, using any technique, sacrifies latency and hence looking at the boundaries of latency that can be offered under certain reliability is of interest to the application in-need of both \cite{8329620}.


%% file: system.tex
\label{sec:system}
In this section, we first introduce the system model for the Cloud-RAN system with multi-path FH and then shed a light on functional split requirements in term of latency.

\subsection{System Model}

Our system model consists of Cloud-RAN with a single CU and a single DU connected with multiple FH paths ($n$ different paths), where each path $i$ has a capacity ${\psi}_{i}$. Packets of size $B$ bits are arrived to the system with exponential inter-arrival periods with average $1/\lambda$ seconds (s). We assume that the FH links are identical. Each link is modelled as a single queue. We suppose that the service time of each queue follows an exponential distribution. The mean service time to transmit a packet of size $B$ bits from CU to DU is $1/\mu = B/\psi$ s. The packets within each queue is served in a first in first out manner and the buffer length is assumed to be infinite. The focus of our model is on downlink (DL) direction. However, all arguments are valid in the reverse direction of communications.

We analyze the performance of the system by considering coexistence of both eMBB and URLLC traffics over orthogonal and non-orthogonal sharing of FH resources as described in \cite{Ghizi} using multi-path FH with coding (MPC).

\begin{figure*}[t]
	\begin{centering}
		\includegraphics[scale=1]{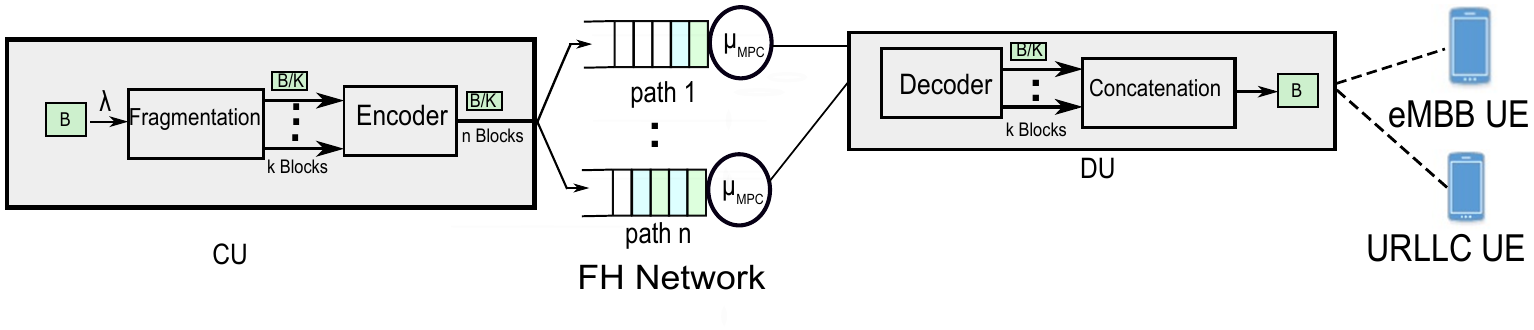}
		\caption{Multi-path FH with erasure coding (MPC) for downlink communication.}   \label{fig:multi_path_fountain}
	\end{centering}
\end{figure*}

In this solution (Fig. \ref{fig:multi_path_fountain}), packets arrive to the CU with exponential inter-arrival periods with average $1/\lambda$ s. Each packet goes through the four steps below,
 \begin{itemize}
 \item \textit{Fragmentation} block fragments the arrival packet into $k$ equal blocks, each with size $B/k$.
 \item \textit{Encoder} block encodes the blocks into $n$ encoded blocks with size $B/k$. Each block is then forked into $n$ paths and serviced in parallel. The service time of each path follows an exponential distribution with service rate $\mu_\textsubscript{MPC} = \frac{k\psi}{B}$.
 \item At the receiver, the original packet can be retrieved if any $k$ out of $n$ are received successfully. Thereby, once $k$ blocks are received, they are passed into \textit {decoder} to start decoding them without waiting for the remaining ones. 
 \item The $k$ decoded blocks are passed to \textit{concatenation} block to be merged into one packet.
 \end{itemize}

Latency in this solution is determined from the time packet is transmitted over the FH until $k$ blocks are successfully received.

\subsection{Functional Split and Latency Requirement}

\begin{table*}[h]
	\caption{Requirements for 5G Scenarios}    \label{table:5G_requirements}
	\begin{center}
		\begin{tabular}{l l l l }
			Scenario & End-to-end latency  & Reliability & Payload size \\
			\hline
			\hline
			Tactile interaction & $0.5$ ms & $99.999\%$ & Small \\
			\hline
			Electricity distribution (high voltage) & $5$ ms & $99.9999\%$ & Small \\
			\hline
			Electricity distribution (medium voltage) & $25$ ms & $99.9\%$ & Small to big \\
			\hline
			Discrete automation & $10$ ms & $99.99\%$ & Small to big \\
			\hline
			Intelligent transport systems & $10$ ms & 99.9999\% & Small to big \\
			\hline
		\end{tabular}
	\end{center}
\end{table*}

\begin{table}[h]
\caption{Bandwidth and latency requirements for different split points}    \label{table:split_requirements}
\begin{center}
\begin{tabular}{l l l l }
Split Point & One-way Latency  & DL Bandwidth & UL Bandwidth \\
\hline
\hline
PDCP-RLC & $1.5-10$ ms & $4016$ Mbps & $3024$ Mbps \\
\hline
MAC-PHY & $250$ $\mu$s & $4133$ Mbps & $5640$ Mbps \\
\hline
\end{tabular}
\end{center}
\end{table}

Different functional split points have different latency and bandwidth requirements on the FH \cite{3gpp.38.801}. These requirements should be considered to support 5G scenarios since each scenario requires different end-to-end requirements in term of latency and reliability as shown in Table \ref{table:5G_requirements} \cite{3gpp.22.261}. Hence, a split point that can sustain scenario requirements can be considered appropriate.

Among all available splits, we will focus our attention on PDCP-RLC and MAC-PHY splits (option 2 and option 6 respectively according to \cite{3gpp.38.801}). The expected latency requirements as estimated by 3GPP \cite{3gpp.38.801} for each split is listed in Table \ref{table:split_requirements}.

 \begin{itemize}
 \item PDCP-RLC split:  For this split, RRC and PDCP are centralized whereas RLC, MAC, PHY and RF are distributed. From a latency point of view, PDCP-RLC split has a relaxed latency requirement on the FH. It tolerates high latency as PDCP doesn't require a strict lower layer synchronization. The maximum tolerable one way latency should be in maximum $10$ ms.
 \item MAC-PHY split: For this option, the split is between MAC and PHY wherein only PHY and RF are in DU. The split offers a high level of centralization and pooling gain compared to PDCP-RLC split. In this split, the HARQ process and other timing critical functions are located in CU which results in tighter latency constraints on the FH. This split can support $250$ $\mu$s latency in maximum.
 \end{itemize}

%% file: bound.tex
\label{sec:bound}


The fronthaul delay for the MPC method described in section \ref{sec:system} can be computed by analysing $(n,k)$ fork-join system. Although $(n,n)$ fork-join system, also known as basic fork-join system, has been thoroughly studied, there are many open problems in analysing its generalization, i.e. $(n,k)$ fork-join system. Mean value analysis for $(n,k)$ fork-join system has been done in \cite{Wang2017} and \cite{Gauri}. Authors in \cite{Fidler2015} used stochastic network calculus to define a stochastic upper bound for distribution of delay in $(n,k)$ fork-join system in a general case. Nevertheless, there is no reasonable way to use that formula without knowing the joint distribution of parallel queues (in case of dependent queues). In this paper, we compute an upper bound and lower bound for $(n,k)$ fork-join system delay using the concept of independency and full dependency between parallel links.

It is worth mentioning that compared to the work presented in \cite{Gauri}, we make an additional assumption of non-purging scenario i.e. after $k$ out of $n$ blocks exit the queuing system, the other $n-k$ remaining blocks are not removed from the queues and they will continue being processed. This assumption is more realistic in this context given dispatched packets can not be removed from the links and switches. As it has been discussed in \cite{Gauri}, split-merge system (which is a variation of $(n,k)$ fork-join system that blocks processing of the next packets until $k$ out of $n$ blocks of the current packet finish being processed) provides an upper bound for delay in purging scenario, however, it is not an upper bound in non-purging scenario. 

In this section, our objective is to compute stochastic delay bounds of $(n,k)$ fork-join system. As detailed in section \ref{sec:system}, in the MPC method, CU encodes the packet into $n$ equal length blocks and sends those blocks into $n$ parallel links. Hence assumption of independence between $n$ links is not valid. Moreover, without knowing the dependency between the links, e.g. their joint distribution, computation of delay bounds are not tractable. Therefore, in this paper we calculate stochastic lower bound and upper bound for delay distribution under certain assumptions. 

\subsection{Stochastic Lower Bound for Delay}

In a homogeneous $(n,k)$ fork-join system, the smallest delay stochastically occurs when all links are independent, since in that case if some links are highly congested, other links might be less congested with larger probability. Therefore, here we will find the stochastic distribution of delay in the case that all links are independent. Authors in \cite{Fidler2015} computed delay bound for independent links, using stochastic network calculus for a general case. However, we will derive this bound for the case in which each parallel link is an $M/M/1$ queue by applying classical queuing theory.

For an $M/M/1$ queue with the iid Poisson arrival process, with mean $\lambda$ and iid exponentially distributed service times with mean $1/\mu$, we have

\begin{equation}
\label{bernoli}
P\{d> \tau \}= e^{-(\mu-\lambda)\tau} =: p_0 \text{ , }
\end{equation}

where $d$ is the block delay in an $M/M/1$ queue which includes waiting time of the block in the queue plus its own service time. Let us assume $(n,k)$ fork-join system, which consists of $n$ parallel homogeneous $M/M/1$ queues. In this system, delay of a packet is defined as the time between execution of the $n$ encoded blocks into the $n$ parallel links until the first $k$ out of $n$ blocks have been processed in the queues. Note that Equation (\ref{bernoli}) can be viewed as a Bernoulli process in which, the number of successes in $n$ independent trials has Binomial distribution. Therefore, $(n,k)$ fork-join delay, denoted by $D$, would be,

\begin{equation}
\label{eq:LB}
P \{D>\tau\} = \sum_{j=0}^{k-1} \binom{n}{j} (1-p_0)^j p_0^{n-j} \text{.}
\end{equation}

which is the probability that more than $n-k$ links have greater delay than $\tau$.

\subsection{Stochastic Upper Bound for Delay}

Similarly in a homogeneous $(n,k)$ fork-join system, the largest delay stochastically occurs when all links are fully dependant, such that all queues have the same length; in this case congestion happens at the same time in all links. Therefore, we use the ``equal queue length'' assumption to compute the worst case of dependency, instead of looking for joint distribution of the queues, and find a stochastic upper bound for $(n,k)$ fork-join system delay. 

For an $M/M/1$ queue, the queue length, denoted by $L_q$, would be equal to $l$ with the following probability,

\begin{align}
P \{L_q=l\} &= (1-\rho)\rho^l \\
\rho &:= \frac{\lambda}{\mu}\text{.}
\end{align}

Also, delay profile for an $M/M/1$ queue with length $l$ is as follows,

\begin{equation}
P \{d>\tau\} = \sum_{m=0}^{l} \frac{(\mu \tau)^m}{m!} e^{-\mu\tau} :=p_1\text{.}
\end{equation}

Similar to the previous computation, to find the delay distribution of $(n,k)$ fork-join system consisting of $n$ homogeneous $M/M/1$ queues, we should compute the probability that more than $n-k$ queues have the delay greater than $\tau$. Therefore, in the case of dependant parallel queues, the $(n,k)$ fork-join system delay will be as follows,

\begin{align}
\label{eq:joint_p}
&P \{D>\tau\} = \sum_{j=0}^{k-1} \binom{n}{j} \times \nonumber \\
&P\{d_1<\tau,..., d_{j}<\tau, d_{j+1}>\tau, ..., d_n>\tau\} \text{,}
\end{align}

where $d_i$, $i=1,..., n$ denotes the block delay in the $i^{th}$ queue (i.e. $i^{th}$ link). In this analysis, we assume all parallel links have the same queue length equal to $L_q$. We further assume that solely this property, i.e. equal queue length, defines the dependency between links, while the queues are assumed to be independent. Hence, joint probability in Eq. (\ref{eq:joint_p}) can be computed using Bayes' law, as follows,

\begin{align}
\label{eq:UB}
&P \{D>\tau\} = \sum_{j=0}^{k-1} \binom{n}{j} \times \nonumber\\
&\sum_{l=0}^{\infty} P\{d_1<\tau,..., d_n>\tau | L_q=l\} P\{L_q=l\}
=\sum_{j=0}^{k-1} \binom{n}{j} \times \nonumber\\
&\sum_{l=0}^{\infty} P\{d_1<\tau| L_q=l\}... P\{d_n>\tau | L_q=l\} P\{L_q=l\} \nonumber
\\
&= \sum_{j=0}^{k-1} \binom{n}{j} \sum_{l=0}^{\infty} (1-p_1)^j p_1^{n-j} (1-\rho) \rho^{l} \text{ . }
\end{align}


%% file: simulation.tex
\label{sec:simulation}

In this section, we develop a simulation model in MATLAB to validate our analysis in the presence of coexisting eMBB and URLLC services. Characterization of the two services are shown in Table \ref{table:SimulationParameters}. We assume there are $n = 10$ independent FH paths, where each path $i$ has a capacity of $100$ Mbps, i.e. ${\psi}_{i} = 100$ Mbps, $\forall i$.

We initially plot the non-orthogonal sharing of FH resources with orthogonal FH transmission schemes that can allocate a different amount of resources to URLLC. In these first plots, the aim is to determine the allocations that improve the probability of error for a given latency for the URLLC services in orthogonal as compared to non-orthogonal FH shared resources.

\begin{table}[h]
\caption{System parameters for the 5G Services}    \label{table:SimulationParameters}
\begin{center}
\begin{tabular}{|p{2.5 cm}|p{1 cm}| p{1 cm}|}
\hline
Type of traffic & eMBB & URLLC \\
\hline
\hline
Packet Size (Bytes) & 1500 & 500 \\
\hline
$\lambda$ (packet/ms) & 4 & 8 \\
\hline
\end{tabular}
\end{center}
\end{table}

\begin{figure}[h]
	\begin{centering}
		\includegraphics[scale=0.27]{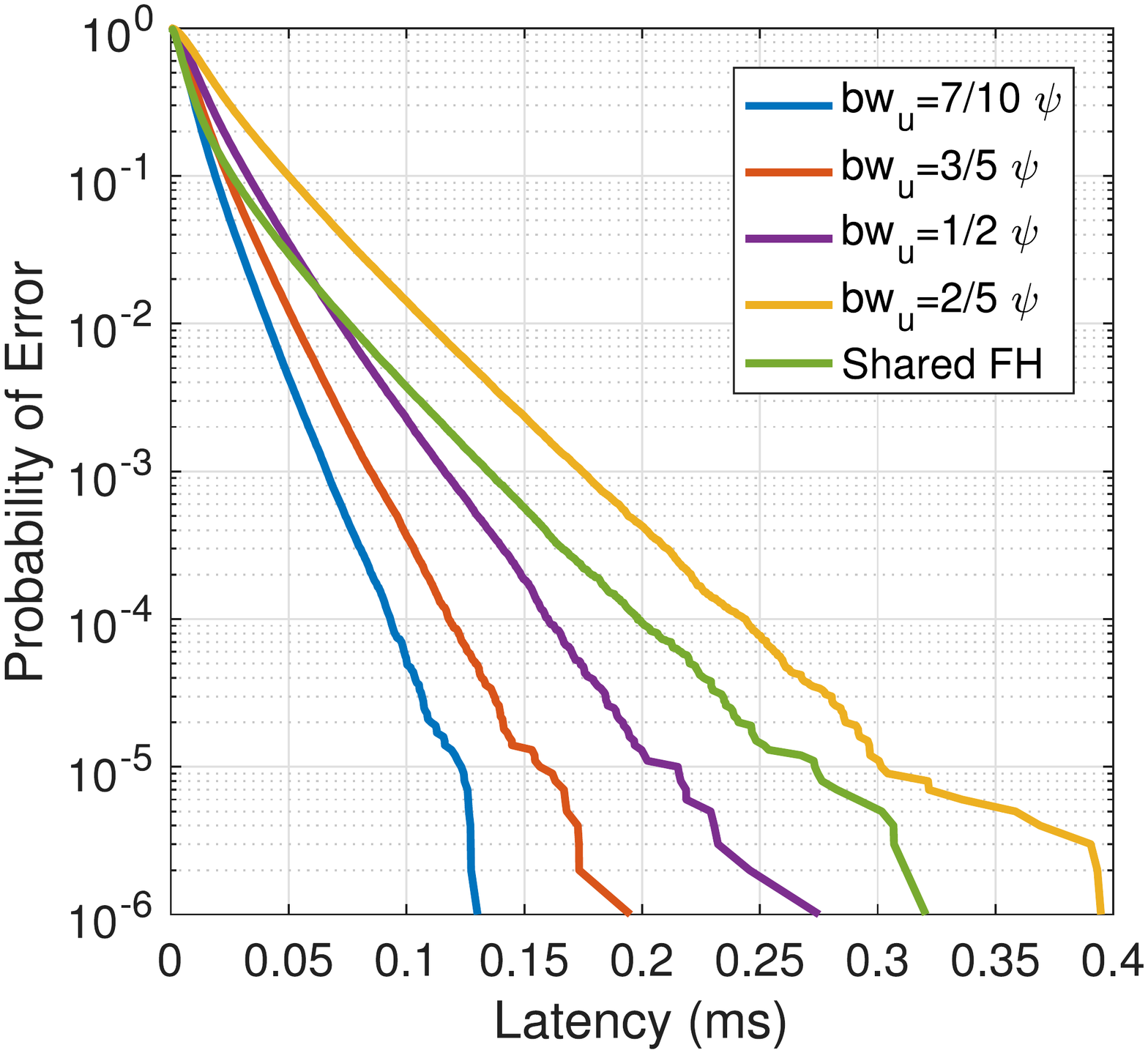}
		\caption{Achievable probability of error Vs. latency under orthogonal FH bandwidth with different bandwidth fractions for URLLC.}   \label{fig:urllc_bw_frac_full_bw}
	\end{centering}
\end{figure}

\begin{figure}[h]
	\begin{centering}
		\includegraphics[scale=0.27]{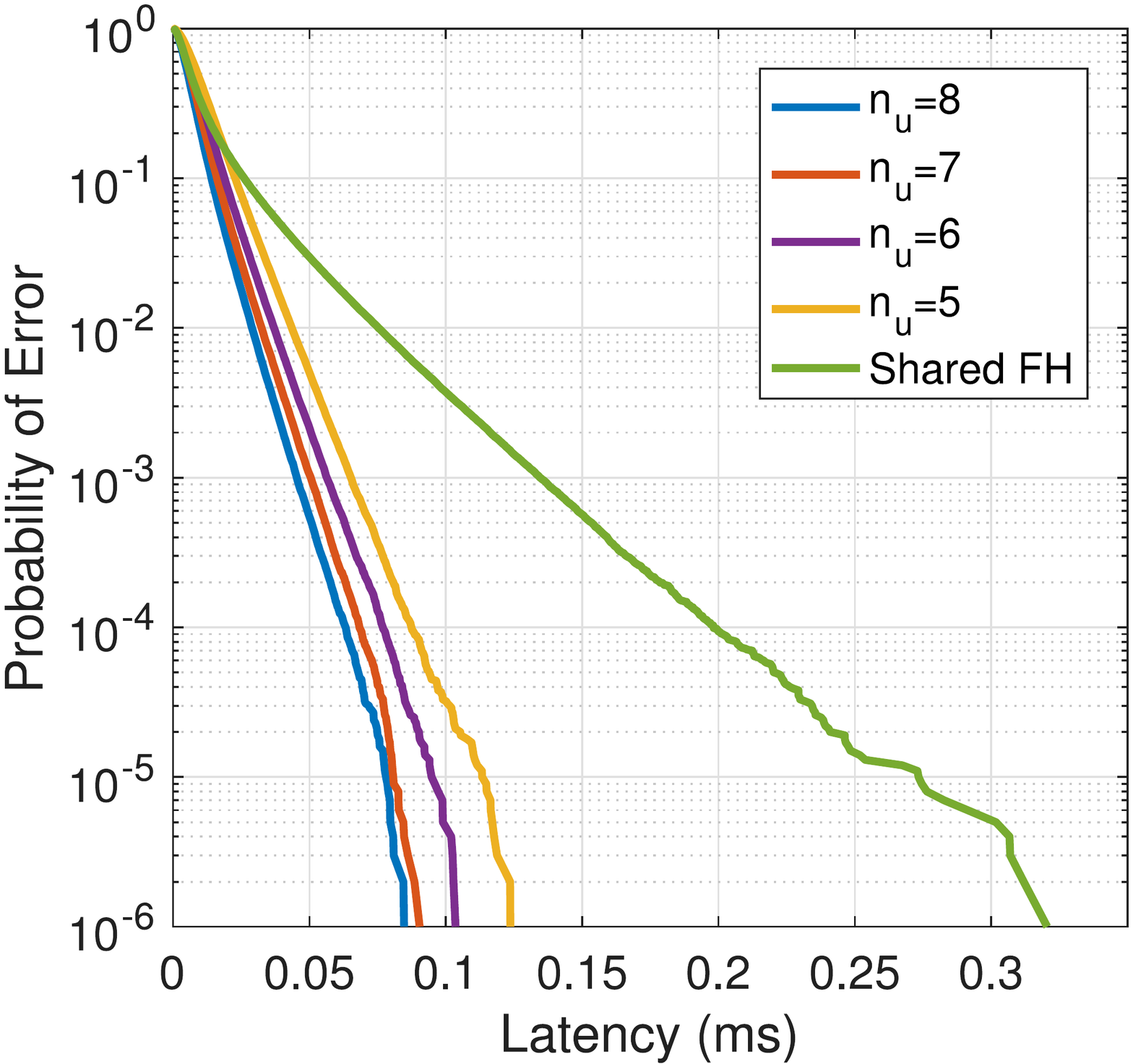}
		\caption{Achievable probability of error Vs. latency under orthogonal FH path with different number of paths for URLLC.}   \label{fig:urllc_path_frac_full_bw}
	\end{centering}
\end{figure}

\begin{figure*}[h]
	\centering
	\subfigure[URLLC.\label{fig:urllc_relia_bw_split_ratio_1_2}]
	{\includegraphics[scale=0.3]{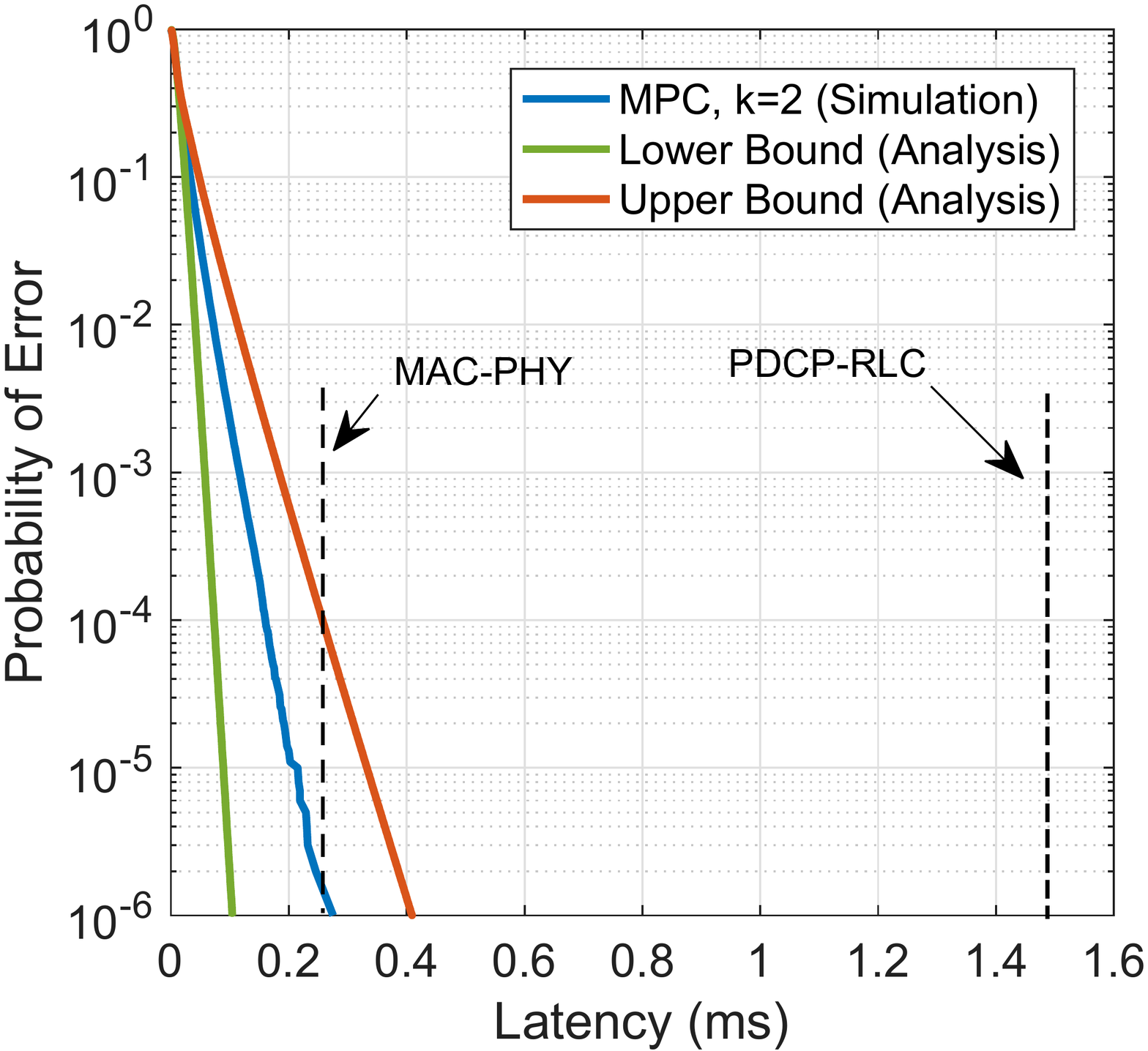}}
	\subfigure[eMBB.\label{fig:embb_relia_bw_split_ratio_1_2}]
	{\includegraphics[scale=0.3]{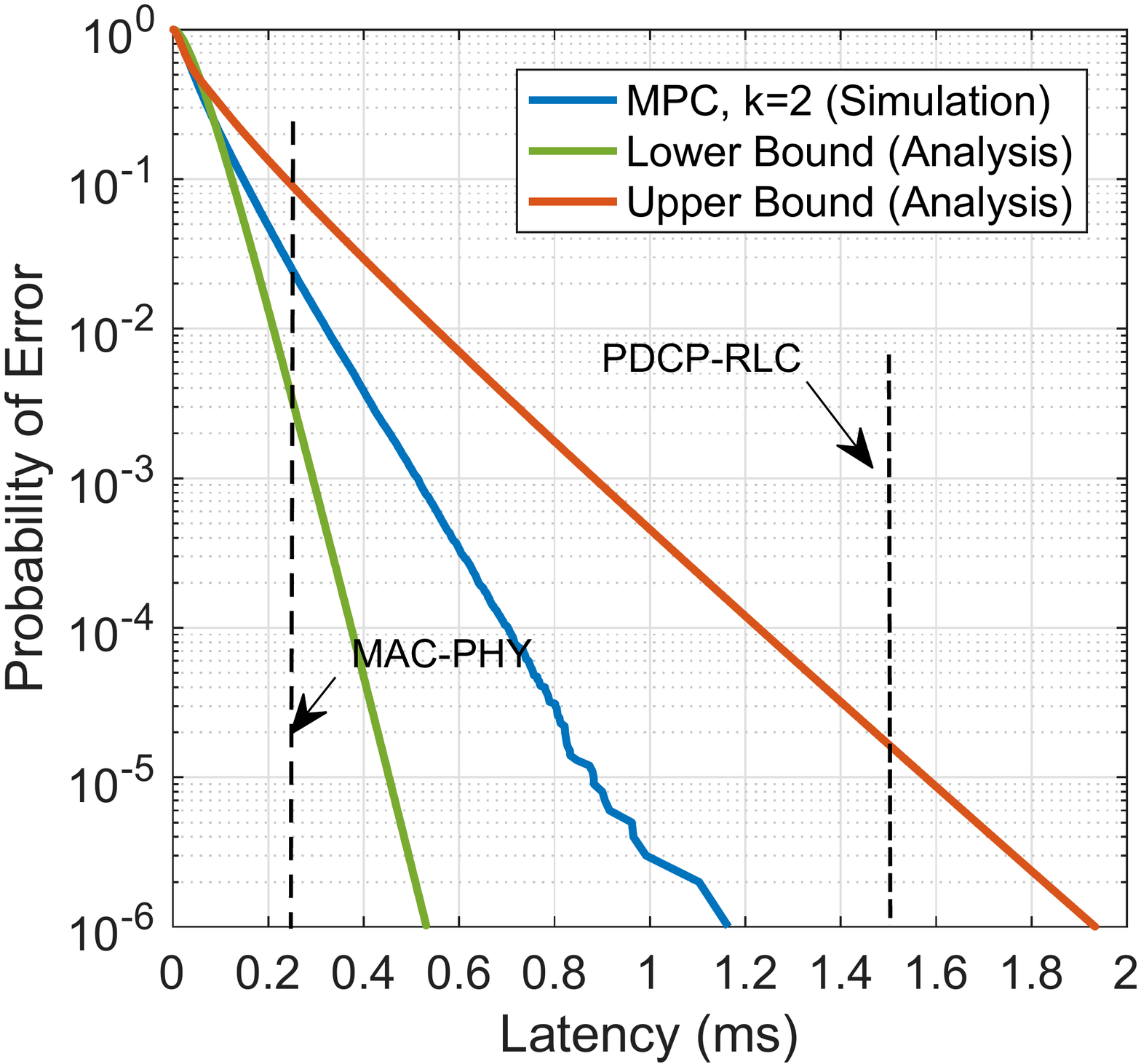}}
	\caption{Achievable probability of error Vs. latency at the upper and lower latency bounds, under bandwidth orthogonal FH with URLLC bandwidth fraction $1/2$ ($bw_e=1/2\psi_{i}$ and $bw_u=1/2\psi_{i}$). \textit{The dashed line represents the maximum latency supported by functional split}}
	\label{fig:relia_bw_split_ratio_1_2}
\end{figure*}

\begin{figure*}[h]
	\centering
	\subfigure[URLLC.\label{fig:urllc_relia_path_split_nu_5}]
	{\includegraphics[scale=0.3]{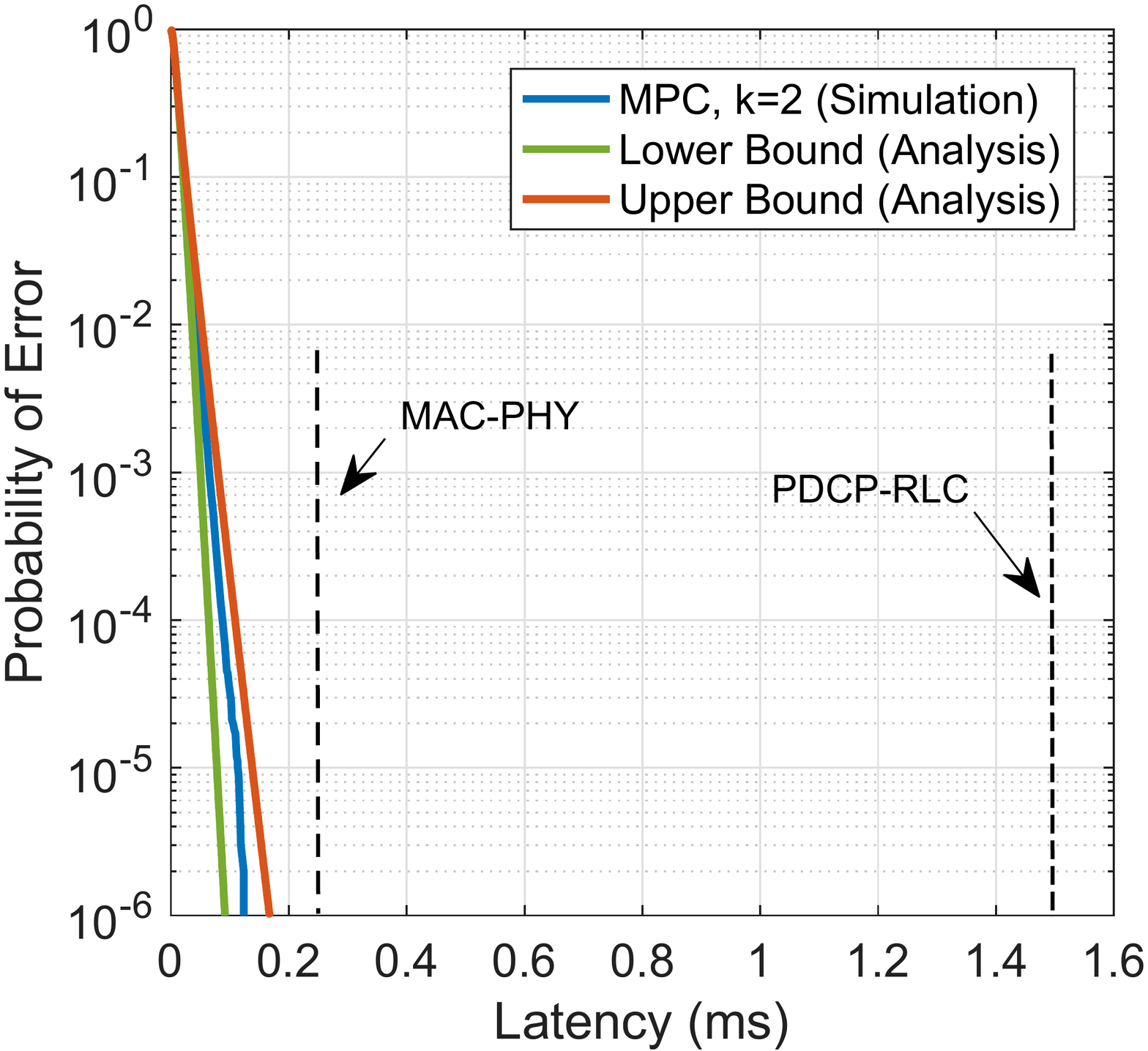}}
	\subfigure[eMBB.\label{fig:embb_relia_path_split_nu_5}]
	{\includegraphics[scale=0.3]{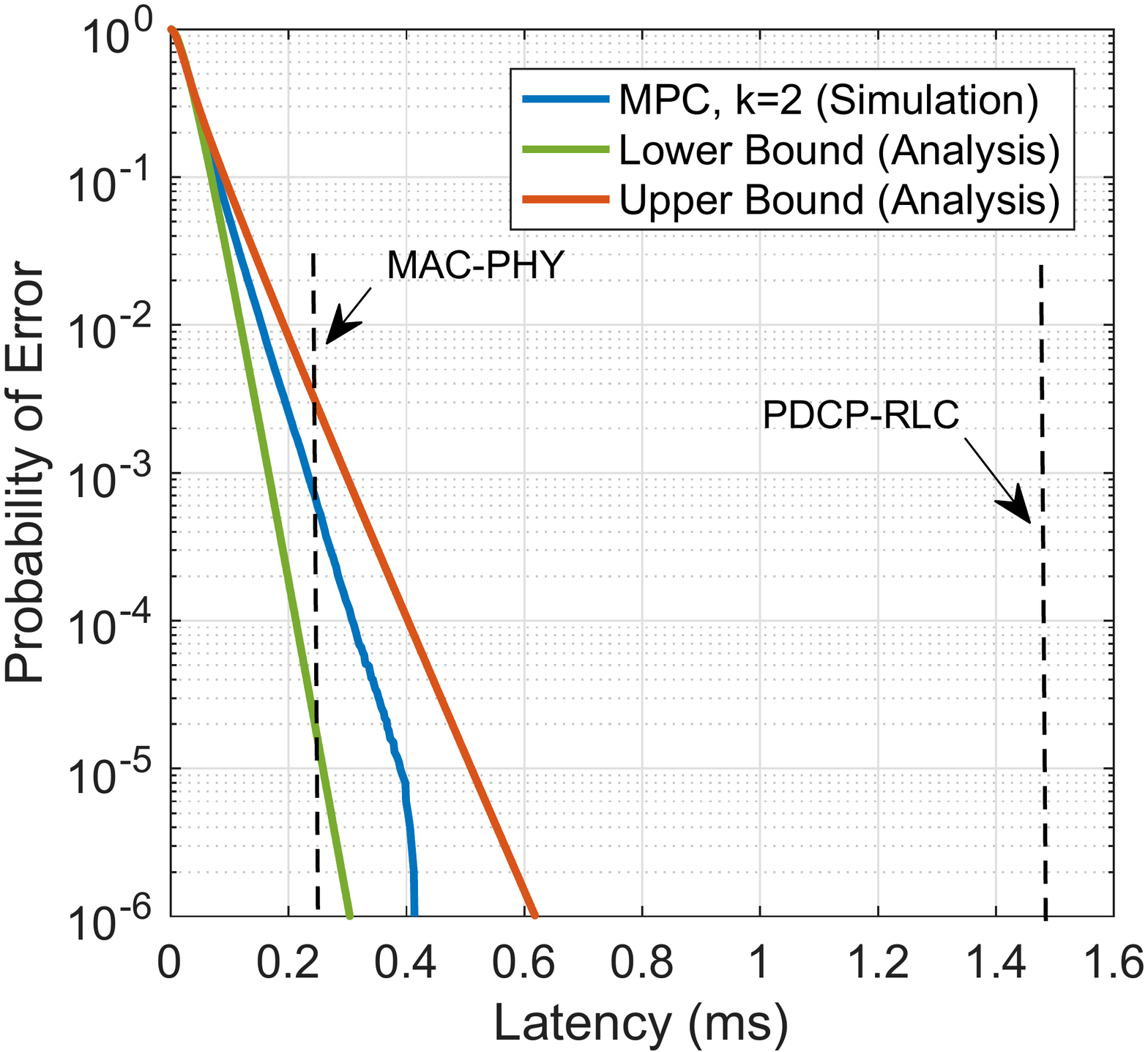}}
	\caption{Achievable probability of error Vs. latency at the upper and lower latency bounds, under path orthogonal FH with URLLC path fraction $1/2$ ($n_e=5$ and $n_u=5$). \textit{The dashed line represents the maximum latency supported by functional split}}
	\label{fig:relia_path_split_nu_5}
\end{figure*}

In Fig. \ref{fig:urllc_bw_frac_full_bw}, we plot the error probability for URLLC using orthogonal bandwidth allocation on the FH with different URLLC bandwidth fractions; in each case $bw_u$ fraction of the available path bandwidth, $\psi_{i}$, is allocated to URLLC. The plot shows that the choice $bw_u \geq 1/2~\psi_{i}$ can reduce the latency as compared to shared FH transport.

Fig. \ref{fig:urllc_path_frac_full_bw} shows the error probability for URLLC using orthogonal path allocation on the FH with different number of paths allocated to URLLC, i.e. using $n_u$ of the available path, $n=10$. The plot shows the latency to achieve the error probability of better than $10^{-1}$ can be reduced as compared to shared FH transport by choosing $n_u \geq 5$. For example the error probability of $10^{-6}$ obtained using orthogonal paths is improved by approximately $160\%$ as compared to that obtained by non-orthogonal sharing of FH resources.

Focusing on orthogonal bandwidth allocation on the FH with $bw_u = 1/2~\psi_{i}$, Fig. \ref{fig:relia_bw_split_ratio_1_2} shows that simulation results are bounded by the lower and upper bounds computed from Equations (\ref{eq:LB}) and (\ref{eq:UB}). For URLLC (Fig. \ref{fig:urllc_relia_bw_split_ratio_1_2}), to achieve a reliability of $99.9999\%$ the latency ranges from $0.1$ ms to $0.4$ ms. As for eMBB (Fig. \ref{fig:embb_relia_bw_split_ratio_1_2}), the latency range is wider varying from $0.53$ ms to $1.92$ ms. From Fig. \ref{fig:relia_path_split_nu_5} we can observe the performance of both URLLC and eMBB are enhanced as compared to orthogonal bandwidth allocation (Fig. \ref{fig:relia_bw_split_ratio_1_2}).

Focusing on Fig. \ref{fig:relia_path_split_nu_5}, we use the lower and upper bounds obtained in this figure to choose appropriate functional split that offers the required reliability for a given scenario. For example using MAC-PHY split with the URLLC traffic (Fig. \ref{fig:urllc_relia_path_split_nu_5}), the upper bound can provide a reliability of $99.9999\%$ at latency of $0.167$ ms. Therefore, this setup can be used for low latency applications which requires a reliability as high as $99.9999\%$. Considering the requirements listed in Table \ref{table:5G_requirements}, this setup can, for example, be used for all scenarios. As for eMBB (Fig. \ref{fig:embb_relia_path_split_nu_5}), MPC with MAC-PHY split can offer a reliability less than $99.9\%$ which is not suitable for any scenario listed in Table \ref{table:5G_requirements} whereby the reliability requirements are of at least $99.9\%$. In such a case MPC with PDCP-RLC split is the only choice available since the upper bound can provide a reliability of $99.9999\%$ at latency of $0.6$ ms.

To summarise, MAC-PHY split is the most appropriate split for scenarios using URLLC traffic considering the system model and traffic patterns in Table III, since it meets their latency and reliability requirements. Whereas, PDCP-RLC split is more suitable for scenarios using eMBB traffic.